\documentstyle[preprint,aps]{revtex}
\begin{document} 
\title{A Linear Approximation for the Excitation Energies of single and 
double analog states in the f$_{7/2}$ shell}
 
\author{Y. Durga Devi, Shadow Robinson and Larry Zamick} 
\address{Department of Physics and Astronomy, Rutgers University, 
Piscataway\\ 
New Jersey 08854-8019, USA} 
\maketitle 
 
\begin{abstract} 
We find that the excitation  energies of single analog states for
odd-even nuclei in the f$_{7/2}$ shell with J=j=7/2$^{-}$  and
the J=0$^{+}$ double analog states in the even-even nuclei 
are well described by the formulas $E^{*}\left(j,T+1\right) = b (T+X)$ and
$E^{*}\left(0^{+},T+2\right) = 2b (T+X+0.5)$,respectively, where $T=\mid
N-Z\mid /2$ is usually the ground state isospin. It is remarkable to
note that the parameter X accounts for the departures from the symmetry
energy based predictions.  
\end{abstract} 
 

\section{Introduction}

In a 1964 Technical Report, McCullen, Bayman and Zamick (M.B.Z) gave 
the wavefunctions and energy levels for nuclei in the f$_{7/2}$ shell 
\cite{MBZ}. In Table 1, we show their results for the excitation energies 
of single analog states in odd-even and even-odd nuclei, with angular momentum
J=j=7/2, also shown are the calculated energies of double analog
states in the even-even Ti isotopes. 

It was noted by M.B.Z \cite{MBZ} that in some cases there was a two to
one relation between the spectra of even-even nuclei and neighbouring
odd A nuclei. For example the calculated J=0$^{+}$ spectra of
$^{44}$Ti were at twice the energies of the corresponding J=j levels
in $^{43}$Ti (or $^{43}$Sc). The same was true for the pairs
$\left(^{48}Ti, ^{47}Sc\right)$, $\left(^{48}Ti, ^{49}Ti\right)$, 
and $\left(^{52}Fe, ^{53}Fe\right)$. That the two to one relation
should hold was easily proved \cite{ZD}. Since for these nuclei it was
true for all levels (of the j$^n$ configuration) which also includes
the double and single analog states. It was noted by Zheng and Zamick 
\cite{ZZ} that the two to one relation holds
quite well experimentally not only for the above mentioned pairs but
for others as well. Zamick and Devi showed \cite{ZD} that the two to
one relation holds approximately for $\left(^{46}Ti, ^{45}Sc\right)$
and the cross conjugate pair $\left(^{50}Cr, ^{51}Cr\right)$. One gets
an exact two to one relation here as well if one excludes 
seniority four states. 

Besides the two to one relation there is the general question of the
systematics of the excitation energies of the single analog states and
double analog states. We will discuss this in the next section.

\section{Empirical Formula for ecitation energies of single and double
analog states in the f$_{7/2}$ shell}

\subsection{Symmetry Energy Formula}

The symmetry energy in the semi-empirical mass formula is
$-a_{SYM}(Z-N)^{2}/A$ with a$_{SYM}$ equal to -21.4 MeV.

Let us define $T=\mid N-Z \mid /2$. For most nuclei T is the ground
state isospin. With this formula the excitation energy of a single
analog state is proprotional to $\left[\left(T+1\right)^2 - T^2\right] =
2(T+1/2)$ whilst the energy of a double analog state is proportional
to $\left[(T+2)^2-T^2\right]=4(T+1)$. More precisely,

\[ E^{*}(S.A.) = - \frac {8a_{SYM}}{A}\left(T+\frac{1}{2}\right) \]
\[ E^{*}(D.A.) = - \frac {16a_{SYM}}{A}\left(T+1\right) \]

\subsection{The multipole interaction $a+bt(1) \cdot t(2)$}

The total potential energy will be $(b/2) T \cdot T + constant$

The S.A. energy will be $(b/2) [(T+1)(T+2)-T(T+1)] = b(T+1)$

The D.A. energy will be $(b/2) [(T+2)(T+3)-T(T+1)] = 2b\left(T+\frac{3}{2}\right)$

\subsection{A Linear Fit}

The previous simple models suggest that the energies are approximately
linear in T (or $\mid N-Z \mid$), we try

\[ E(S.A.) = b (T+X)\]

\[ E(D.A.) = 2b (T+X+1/2)\]

In Table 1 we fit the above formulas to the calculated energies of 
single j shell calculations in which the spectrum of $^{42}$Sc was used
to determine the matrix elements $\left<\left(j^2\right)^I
\mid V \mid \left(j^2\right)^I\right>$, $j=f_{7/2}$ and I ranges from
zero to seven. The matrix elements, in MeV, are 0, 0.6111,
1.5863, 1.4904, 2.8153, 1.5101, 3.242 and 0.6163, respectively. 

We find that we get a good fit with b=2.32 MeV and X=1.30. The formulas
do not give an exact fit, but the results are nevertheless very good. 
For this linear fit there are several results which are independent of 
the values of the parameters b and X. For example, states with the same 
T should have the same S.A. excitation energies in the shell model 
calculation. However, the single j shell results are very close 4.142 MeV 
and 4.112 MeV. Wherever the single j shell gives a two to one ratio for 
the energies of D.A. states as compared with S.A. states. So does the 
linear fit, irrespective of what b and X are. 

The single j shell calculation does give two to one ratios of D.A. to
S.A. for the pairs $\left(^{44}Ti, ^{43}Ti\right)$, $\left(^{44}Ti, 
^{43}Sc\right)$, $\left(^{48}Ti, ^{49}Ti\right)$ and $\left(^{48}Ti, 
^{47}Sc\right)$, $\left(^{52}Fe, ^{53}Fe\right)$ and so also do the linear
formulas. But the linear formulas also give two to one ratios where the
single j shell does not. These include $\left(^{46}Ti, ^{45}Sc\right)$ 
and $\left(^{46}Ti, ^{47}Ti\right)$. In the single j shell the
excitation energies are 13.204 MeV for $^{46}Ti$ and 6.590 MeV for
$^{47}Ti$ yielding a ratio 2.0036.

The $\left(^{46}Ti, ^{45}Sc\right)$ case was considered by Zamick and
Devi \cite{ZD}. They noted that if in the single j shell calculation one
neglected seniority four admixtures then one would get a two to one
relation because then the dimensions of the basis states would be the
same --- four (In the exact case they were 6 for $^{46}$Ti and 7 for
$^{45}$Sc). Indeed in the 4 $\times$ 4 diagonalization there will be a
two to one relation for $\underline{all}$ the states, not just the
analog state and as mentioned in ref.\cite{ZD} even if one does not
neglect seniority 4 states one can see by eyeball an approximate
correction between the energies and wavefunctions of several of the
states in the two nuclei. 

However, the case $\left(^{46}Ti, ^{47}Ti\right)$ is different. 
There are 17 J=j basis states for $^{47}$Ti, but as previously
mentioned, only, six for $^{46}$Ti. Nevertheless the D.A. analog
excitation energy in the single j shell calculation is very close to
twice that of the S.A. excitation energy in $^{47}$Ti. The actual
ratio is 2.0036.

Note also that the S.A. states in $^{45}$Sc and $^{47}$Ti have nearly
the same excitation energies. Again the configurations look completely
different. In $^{45}$Sc we have one proton and four neutrons. The four
neutrons could have seniority 0,2 or 4. In $^{47}$Ti we have 2 protons
and three neutron holes. The only common thread between the two nuclei
is that they have the same neutron excess N-Z=3. Indeed the linear
approximation yields both the two to one relation between the D.A. in
$^{46}$Ti and the S.A. in $^{47}$Ti and and the equality of the S.A. 
excitation energies of $^{45}$Sc and $^{47}$Ti.

\section{The Experimental Situation}

Previously Zamick and Zeng \cite{ZZ} made a comparison of single analog
and double analog excitation energies with shell model calculations.
While the main thrust of the present work is to discuss surprizing near
degeneracies of SA and DA excitations in the single j shell model, a
reexamination of the experimental situation would certainly be of
interest and of value. 

We will here make a fit to the D.A. excitations in the Titanium isotopes
with our linear formula E$^{*}$(D.A.)=2b(T+X+0.5). The results are as
follows:

\begin{tabular*}{6in}{@{\extracolsep{\fill}}ccc}
Nucleus & E$^{*}$(D.A.) (MeV) & 2b(T+X+0.5) \\
 & Experiment & X=2.00 b=1.94 MeV \\
$^{44}$Ti & 9.340 & 9.70 \\
$^{46}$Ti & 14.153 & 13.58 \\
$^{48}$Ti & 17.379 & 17.46 
\end{tabular*}

The values of X and b in Table. 1 are quite different (X=1.30, b=2.32).
To get better agreement will require more extensive calculations
involving configuration mixing, Coulomb interactions and A dependence of
the parameters. Concerning the latter point it should be noted that in
$^{52}$Fe, the value of E$^{*}$(D.A.) is 8.559 MeV. With fixed b,X (or
in a single j shell calculation with fixed two body matrix elements)
this energy should be the same as that in $^{44}$Ti. We can parametrize
the difference by assuming that b varies as $1/\sqrt{A}$. This is
different from the 1/A dependence of the previously mentioned symmetry
energy in the semi-empirical mass formula. 

In the previous references the $\underline{experimental}$ values for
several S.A. and D.A. excitation energies were given. We will here
give these and a few more, grouped according to the isospin. For half
integer isospin we give the S.A. excitation energies while for integer
isospin we give the D.A. excitation energies. The units are MeV. We
also list nuclei for which the energies are not known followed by a
question mark. Hopefully this will stimulate experimental activity. 

\begin{tabular*}{6in}{@{\extracolsep{\fill}}ll}
T=0 & $^{44}$Ti (9.340), $^{48}$Cr (8.75), $^{52}$Fe (8.559) \\
T=1/2 & $^{43}$Sc (?), $^{43}$Ti (?), $^{45}$Ti (4.176), $^{49}$Cr
(4.49), $^{51}$Mn (4.451), $^{53}$Co (4.390), $^{53}$Fe (4.250) \\
T=1 & $^{46}$Ti (14.153), $^{50}$Cr (13.222) \\
T=3/2 & $^{45}$Sc (?), $^{47}$Ti (7.187), $^{51}$Cr (6.611) \\
T=2 & $^{48}$Ti (17.379) \\
T=5/2 & $^{47}$Sc (?), $^{49}$Ti (8.724) 
\end{tabular*} 

\section{Conclusions}

We find that we can give a very good but not perfect fit to a wide
variety of single analog and double analog excitation energies by using
the formulas
$E^{*}(S.A.)=b(T+X)$ and $E^{*}(D.A.)=b\left(T+X+\frac{1}{2}\right)$
where $T=\mid N-Z\mid/2$. The value of X is however different from that
given by simple symmetry energy arguments. For fixed b the formulas
predict that states with the same T will have the same excitation
energies. In the single j shell calculation this is true in some special
cases e.g. for cross conjugate pairs, but is not true in general. But
even when not true the excitation energies are remarkably similar e.g.
$^{45}$Sc, $^{47}$Ti. They differ by only 10 keV in the single j shell
calculation. 

The above formulas give two to one ratios for the
excitation energies of D.A. states of isospin T (integer) and S.A.
states of isospin $T+\frac{1}{2}$. In some special cases the single j
shell calculation also gives the result e.g.
$\left(^{44}Ti,^{43}Ti\right)$. In other cases it does not but
nevertheless the results are very close e.g. $^{50}Cr, ^{51}Cr$. The
$\left(^{50}Cr, ^{51}Cr\right)$ case can be explained by seniority
truncation (i.e. neglecting seniority 4 admixtures) but the closeness of
the single analog excitation in $^{45}$Sc and $^{47}$Ti cannot. 
In the former case the components of the wavefunctions look visually
similar, but for $^{45}$Sc and $^{47}$Ti they look completely different.

A look at the experimental data shows that the grouping of excitation
energies according to isospin T (or neutron excess (N-Z)) is quite
appropriate, although it would appear that the parameter b should have
an A dependence. A rough analysis suggests that b is proportional to
$1/\sqrt{A}$.

In the future it will be of interest to see if we can get a better
understanding of the parameter X i.e. why it differs from
$\underline{unity}$. Also the effect of coulomb energies and
configuration mixing, which was partly dealt with in ref. \cite{ZZ}
could be extended. 

\vskip 1cm 
{\bf Acknowledgements}

This work was supported by the Department of Energy Grant No.DE-FG02-95ER40940.

\begin{table}
\caption{The Excitation Energies of Single Analog (J=j) states (S.A.)
and Double Analog (J=0$^+$)
States (D.A.). A comparison is made of single j shell calculations
using the spectrum of $^{42}$Sc as input and linear fits.}
\begin{tabular}{cccc}
Single Analog & Single j & b(T+X)\tablenotemark[1] & Formula \\
 & MeV & MeV & \\
$^{43}Ti\left(^{53}Co\right)$\tablenotemark[2] & 4.142 & 4.176 & b(0.5+X)  \\
$^{45}Ti$ & 4.112 & 4.176 & b(0.5+X)  \\
$^{45}Sc\left(^{51}Cr\right)$ & 6.601 & 6.496 & b(1.5+X)  \\
$^{47}Ti\left(^{49}V\right)$ & 6.590 & 6.496 & b(1.5+X)  \\
$^{49}Ti\left(^{47}Sc\right)$ & 8.829 & 8.816 & b(2.5+X)  \\
 & & & \\
Double Analog &  & 2b(T+X+1/2)\tablenotemark[1] & \\
$^{44}Ti\left(^{52}Fe\right)$ & 8.284 & 8.352 & 2b(0.5+X)  \\
$^{48}Cr$ & 8.000 & 8.352 & 2b(0.5+X)  \\
$^{46}Ti\left(^{50}Cr\right)$ & 13.204 & 12.992 & 2b(1.5+X)  \\
$^{48}Ti$ & 17.659 & 17.632 & 2b(2.5+X)  \\
\end{tabular}
\tablenotetext[1]{$b=2.32 MeV\;\;\;\;X=1.30 \\ T=\mid N-Z \mid /2$}
\tablenotetext[2]{Also the Mirror Nuclei $^{43}Sc\left(^{53}Fe\right)$}
\end{table}


\begin{references} 
\bibitem{MBZ}  J.D. McCullen, B.F. Bayman and L. Zamick, Phys. 
Rev. {\bf 134B} (1964); Technical Report {\bf NYO-9891}.
 
\bibitem{ZD}  L. Zamick and Y.D. Devi, Phys. Rev. {\bf C} (1999) to
appear in print.

\bibitem{ZZ}  L. Zamick and D.C. Zheng, Phys. Rev. {\bf C46} (1992) 815.

\end{references}
\end{document}